\documentclass[12pt]{elsarticle}

\usepackage{subcaption}
\usepackage[utf8]{inputenc}
\usepackage[export]{adjustbox}
\usepackage{wrapfig}
\usepackage{graphicx}
\usepackage{amssymb}
\usepackage{amsmath, nccmath}
\usepackage{commath}
\usepackage{amsthm}
\usepackage{geometry}
\usepackage{lineno}

\usepackage{lipsum}
\makeatletter
\def\ps@pprintTitle{%
 \let\@oddhead\@empty
 \let\@evenhead\@empty
 \def\@oddfoot{}%
 \let\@evenfoot\@oddfoot}
\makeatother

\begin{document}

\begin{frontmatter}


\title{Casimir-like effect from thermal field fluctuations}



\author{Venkat Abhignan*, R. Sankaranarayanan}

\address{Department of Physics, National Institute of Technology, Tiruchirappali - 620015, India. \\ $^*$Email address: yvabhignan@gmail.com}

\begin{abstract}
Landau-Ginzburg $\phi^4$ field theory is usually applied to systems for understanding continuous phase transitions at critical points. Here we analyze thermal field using similar free energy description in statistical field theory perspective, and study fluctuations in such a field with particular focus on realizing thermal Casimir effect. Initially we qualitatively describe emergence of Casimir-like effect using mean-field approximation and further derive it using coarse-graining of perturbative renormalization procedure in the vicinity of Gaussian-fixed point. These results may lead to further the understanding of Casimir effect from scalar fields without employing the concept of zero-point energy in a cosmological sense.
\end{abstract}


\end{frontmatter}


\section{Introduction}
Casimir effect was initially accounted for an interaction between two metal plates \cite{Casimir:1948dh,Bordag:2009zz}. This was observed by measuring change in electromagnetic zero-point energy (ZPE) by changing the distance between plates. Casimir interpreted that electromagnetic fluctuations whose wavelength are comparable with the distance between plates would be contributing to Casimir effect, and this interaction is independent of material in the plates. This result is universal as it depends only on Planck constant $h$, speed of light $c$ and distance between the plates. Due to presence of $h$, Casimir effect is sometimes interpreted as a manifestation of non-trivial quantum vacuum. Planck initially introduced the possibility of ZPE in reference to quantum vacuum, however he found no physical significance to it \cite{Milonni}. Consequently Casimir effect was also realized without reference to ZPE in significant works of varying background potentials \cite{PhysRev.73.360,Lifshitz:1956zz,SCHWINGER19781,RevModPhys.81.1827,zpe1} and related toy models \cite{zpe2,zpe3}. Importance of vacuum energy in description of Casimir effect remains controversial \cite{article1,article2}. \\ Later studies revealed that any dynamical phenomenon such as field interaction with boundaries, non-trivial topology, space-time curvature can be related to Casimir effect at zero-temperature limit. Further, when vacuum state is considered at finite temperature, thermal fluctuation of electromagnetic field becomes an important component of  Casimir effect. This thermal Casimir effect was extensively studied initially by Brown and Maclay at finite temperature \cite{theory0}, followed by recent experimental verification \cite{nature0}. Regarding ZPE in a cosmological aspect, thermal Casimir effect can be considered as effect due to polarization of vacuum between the plates. This polarization energy of scalar fields in topologically non-trivial space-time is attributed to identification conditions and not due to boundary conditions in case of absence of plates \cite{DEWITT1975,QVEGR}. Studies on cosmological Casimir effect particularly in  $S^3 \times R^1$ static space (Einstein and Friedmann universes) led to computation of non zero-temperature contributions to zero-temperature Casimir stress-energy tensor of the scalar field which showed $\sim(k_B T^4)$ behaviour at high-temperature limit, where $k_B$ is Boltzmann constant and $T$ is temperature \cite{vstft}. Similar results were reproduced at low T limit as well \cite{Bezerra2011}. \\ Further critical Casimir effect was formulated and extensively studied in the theory of critical phenomenon at vicinity of second-order phase transitions with boundary conditions \cite{theory1,revmodphys,theory2,nature1}. Landau-Ginzburg $\phi^4$ field theory is considered universal \cite{1967RPPh...30..615F,kadanoff} in sense that different types of continuous phase transitions \cite{Landau:1937obd} have same statistical description of their order parameter. Well known examples are density field in liquid-gas phase transition, magnetisation field in ferromagnetic-paramagnetic phase transition, two component complex field in superfluid transition and polarization field in ferroelectric-paraelectric phase transition. Statistical properties of fluctuations in these different fields are considered to be similar around criticality for a wide range of physical systems \cite{kardar2007statistical}, which can be described by the universal critical exponents from $O(n)$-symmetric $\phi^4$ field theory \cite{Shalaby_2020,abhignan2020continued}. Temperature is a common governing parameter in all these fields which drives the system towards criticality. Entire statistical description of a system can be captured by free energy as a function of temperature and volume. \\ Here we qualitatively attempt to study thermal field as a similar statistical scalar field in $O(1)$ $\phi^4$ model, by considering correlation length of thermal field fluctuations comparable with distance between plates in thermal Casimir effect. This correlation length tends to infinity at criticality, which cannot be related to material bodies but rather can have a cosmological significance \cite{Bezerra2011}.  Analogous to our approach, hypothetical scalar fields were recently studied in this Landau-Ginzburg framework to resolve cosmological issues since microscopic symmetry (translational and rotational symmetries) in $\phi^4$ model can be related to large scale homogeneity and isotropy in a Friedmann universe \cite{GLDE}.

\section{Temperature as a scalar field}

Let us take Landau-Ginzburg kind of Hamiltonian \cite{kardar2007statistical} for a scalar thermal field $T(\mathbf{x})$ in $d$ dimensions at zero-temperature limit. Further here $T(\mathbf{x})$ indicates coarse-grained scalar temperature field over some intrinsic length scale, whereas $T$ indicates temperature determined experimentally. Since temperature is small, ignoring higher order terms and assuming microscopic symmetry we have scaled free energy of system as 
\begin{equation}
    \Phi(T(\mathbf{x})) = \int d^d \mathbf{x} \left[ \frac{t}{2} T^2 (\mathbf{x}) + u T^4 (\mathbf{x}) + \frac{K}{2}(\nabla T (\mathbf{x}))^2\right], \label{1}
\end{equation}
 where $t$, $u$ and $K$ are analytical functions of temperature and can be expanded around mean value of $T(\mathbf{x})$ ($\overline{T}$ is ensemble average of $T(\mathbf{x})$) in Taylor series such as  \begin{align}
 t(T)=t_0+t_1 (T-\overline{T})+t_2 (T-\overline{T})^2+\cdots,
     \\ K(T)=K_0+K_1 (T-\overline{T})+K_2 (T-\overline{T})^2+ \cdots 
     \\ \hbox{and}\,\,\,\,u(T)=u_0+u_1 (T-\overline{T})+u_2 (T-\overline{T})^2+ \cdots .
     \end{align}
Corresponding partition function of system is
 \begin{equation}
    Z = \int DT(\mathbf{x}) \exp\{-\Phi(T(\mathbf{x}))\}, \label{2}
 \end{equation}
 where $\int DT(\mathbf{x})$ indicates integrating over all allowed configurations of the field. Using saddle point approximation we can argue that integral can be evaluated just at point where the integrand has its maximum value, under thermodynamic limit $\int d^d \mathbf{x} \rightarrow \infty $. In the integrand of eq.(\ref{1}), we impose an analogue of thermodynamic equilibrium by taking the term $\frac{K}{2}(\nabla T)^2$ to be maximum to maintain a uniform temperature across all dimensions of space. This amounts to minimization of $f(T)=\frac{t}{2} T^2 (\mathbf{x}) + u T^4 (\mathbf{x})$, which leads to restriction in integration of subspace of $Z$ in eq.(\ref{2}) and saddle point free energy is
 \begin{equation}
     \Phi_{sp} = -\ln Z_{sp} \approx V \hbox{min} \{f(T(\mathbf{x}))\}_T,
 \end{equation}
 where $V$ is volume in $d$ dimensional space. Average temperature $\overline{T}$ in this space around which fluctuations may happen can be given by
 \begin{equation}
    f'(\overline{T}) = t \overline{T} + 4 u \overline{T}^3 = 0, \label{3} \\
 \end{equation}
  \begin{equation}
\overline{T} = \Bigg\{\begin{array}{l}
  \ \ \ \ \ \ \ 0 \ \ \ \hbox{for}\ t>0 \\ 
   \pm \sqrt{\frac{-t}{4u}} \ \ \ \hbox{for}\ t<0.
  \end{array} \label{tbar} 
  \end{equation}
  Here negative solution of temperature is ignored since it provides no physical significance.

 \section{Thermal Fluctuations}
 Let us consider a small non-uniform function $\phi(\mathbf{x})$ to uniform mean temperature $\overline{T}$ all over field $T(\mathbf{x})$ as
 \[T(\mathbf{x})=\overline{T}+\phi(\mathbf{x})\]
 We have \begin{equation}
Z = \exp\left\{-V\left[\frac{t}{2}\overline{T}^2+u\overline{T}^4\right]\right\}\int DT(\mathbf{x}) \exp\left\{-\frac{K}{2}\int d^d \mathbf{x} \left[ (\nabla\phi)^2+\left(\frac{t + 12u\overline{T}^2}{K}\right)\phi^2 \right]\right\}.
 \end{equation}
 By taking fourier modes of thermal fluctuation $$\phi(\mathbf{x})=\sum_{\mathbf{q}} \frac{1}{\sqrt{V}}\phi_\mathbf{q} e^{i\mathbf{q}.\mathbf{x}},$$ we evaluate different terms in partition function as follows;
 \begin{multline}
     \int d^d \mathbf{x} (\nabla\phi)^2 = \int d^d \mathbf{x} \nabla\left( \sum_{\mathbf{q}} \frac{1}{\sqrt{V}}\phi_\mathbf{q} e^{i\mathbf{q}.\mathbf{x}} \right) \nabla\left( \sum_{\mathbf{q}'} \frac{1}{\sqrt{V}}\phi_\mathbf{q}' e^{i\mathbf{q}'.\mathbf{x}} \right) \\ 
     = \frac{1}{V}\sum_{\mathbf{q}} \sum_{\mathbf{q}'} \phi_\mathbf{q} \phi_{\mathbf{q}'}(-\mathbf{q}\mathbf{q}')\int d^d \mathbf{x}  \left( e^{i(\mathbf{q}+\mathbf{q}').\mathbf{x}} \right).
  \end{multline}
  \begin{equation} \hbox{ Since}\,\,
   \int d^d \mathbf{x} \left(e^{i(\mathbf{q}+\mathbf{q}').\mathbf{x}}\right) = V\delta_{\mathbf{q}+\mathbf{q}',0} \,\,\hbox{,}\,\, \int d^d \mathbf{x} (\nabla\phi)^2 = \sum_{\mathbf{q}} q^2\abs{\phi_\mathbf{q}^2}.   \end{equation} 
   \begin{equation}
       \hbox{Similarly} \int d^d \mathbf{x} \phi^2 = \sum_{\mathbf{q}} \abs{\phi_\mathbf{q}^2} .  
   \end{equation}
With this partition function of Landau-Ginzburg Hamiltonian becomes
\begin{equation}
Z = \exp\left\{-V\left[\frac{t}{2}\overline{T}^2+u\overline{T}^4\right]\right\}\int d\phi_\mathbf{q} \exp\left\{-\frac{K}{2}\sum_{\mathbf{q}} \abs{\phi_\mathbf{q}^2}\left[ q^2+\frac{1}{\xi_l^2} \right]\right\},
 \end{equation} where $\xi_l$ is a characteristic length scale and is defined as \[\xi_l^2=\frac{K}{t + 12u\overline{T}^2} \,\,. \]   From partition function of Eq. (13) we can observe that fluctuation for each mode behaves like a Gaussian random variable with mean zero and two-point correlation functions are
  \begin{equation}
     \left\langle \phi_\mathbf{q} \phi_{\mathbf{q}'}\right\rangle = \frac{\delta_{\mathbf{q}+\mathbf{q}',0}}{K(q^2+\xi_l^{-2})}. \label{11}
 \end{equation} 
 Probability of each mode can be defined by integrating over orthogonal Gaussian modes as \begin{equation} 
Z = \exp\left\{-V\left[\frac{t}{2}\overline{T}^2+u\overline{T}^4\right]\right\}\prod_{\mathbf{q}} \left( \frac{2\pi}{K(q^2+\xi_l^{-2})}\right)^\frac{1}{2} . \label{12}
 \end{equation}
With $\Phi = -\ln Z$, energy of system is given by
 \begin{equation}
     E = T^2\frac{\partial \ln Z}{\partial T}. \end{equation}
 From eq. (2) we have
 \begin{equation}
      dt=t_1 dT+2t_2(T-\overline{T})dT+\cdots, \label{t1}
 \end{equation}
 and neglecting higher order terms, average energy is approximated as 
 \begin{equation}
     E \approx T^2(t_1+2 t_2 (T-\overline{T}))\frac{\partial \ln Z}{\partial t}. \end{equation}
     In limit of thermodynamic equilibrium ($T \rightarrow \overline{T}$), we have 
   \begin{equation}
       E\approx t_1 \overline{T}^2 \frac{\partial \ln Z}{\partial t}. \label{16}
   \end{equation}  
   \subsection{Correlation length $\xi_l$}

 From eq.(\ref{11}) we observe that parameter $1/\xi_l$ decides modes of fluctuations where fluctuations in $ \left< \abs{\phi_\mathbf{q}}^2\right> $ are significant, beyond which it decays as a power law. In other words, modes of fluctuations ranging from zero to $1/\xi_l$ are dominant. Also when we derive fluctuation correlations in real space $\left< (\phi(\mathbf{x})-\phi(\mathbf{x}'))^2\right>$ \cite{kardar2007statistical}, we can deduce that range of these fluctuations is $\xi_l$, beyond which fluctuations decay off. Hence this characteristic length scale $\xi_l$ is also called as correlation length.
   \subsection{$\overline{T}\neq 0$}
   From eq.(\ref{tbar}) $\overline{T}=0$ for $t>0$, but since we are interested in studying temperatures close to absolute zero and fluctuations around them we shall consider region $t<0$ where $\overline{T}=\sqrt{\frac{-t}{4u}}$ and obtain
   \begin{equation}
      \frac{K}{\xi_l^2}=-2t=8u\overline{T}^2.
   \end{equation}
   This implies that correlation length of temperature fluctuations is inversely proportional to corresponding temperature. In other words considering eqs. (3) and (4) in the limit $T \rightarrow \overline{T}$, ratio $ K/u \approx K_0/u_0 = 8\xi_l^2\overline{T}^2$, a constant.
     With this, correlation length becomes
   \begin{equation}
       \xi_l=\frac{\sqrt{K/8u}}{\overline{T}}. \label{xi}
   \end{equation}
   Further finding energy of such a system using eqs.(20), (\ref{16}) and (\ref{12}), \[
  \ \ \ \ \ \ \ \ E=t_1 \overline{T}^2 \frac{\partial}{\partial t} \ln \left(\exp\left\{-V\left[\frac{-t^2}{16u}\right]\right\}\prod_{\mathbf{q}} \left( \frac{2\pi}{K q^2-2t}\right)^\frac{1}{2}\right) 
 \]
 \[ =t_1 \overline{T}^2 \left(\frac{Vt}{8u} - \frac{1}{2} \sum_{\mathbf{q}} \frac{\partial}{\partial t} \ln(K q^2-2t)\right) .\]  Taking a continuous spectrum for modes $\mathbf{q}$, energy density is \[
    \frac{E}{V}=t_1 \left(\frac{-t}{4u}\right) \left(\frac{t}{8u} + \int \frac{d^d\mathbf{q}}{(2\pi)^d} \frac{1}{(K q^2-2t)}\right).
 \] Further implementing eq.(20) and the magnitude of thermal fluctuations given by $\left\langle \abs{\phi_\mathbf{q}}^2 \right\rangle =\frac{1}{K(q^2+\xi_l^{-2})}$, \begin{equation}
      \frac{E}{V}= \frac{-t_1K^2}{128u^2\xi_l^{4}} + \frac{t_1K}{8u\xi_l^{2}}\int \frac{d^d\mathbf{q}}{(2\pi)^d} \left\langle \abs{\phi_\mathbf{q}}^2 \right\rangle . \label{meane}
 \end{equation} 
  
 \section{Thermal Casimir-like force}
      Wien's displacement law states that wavelength $\lambda_{max}$ at which energy density peaks for electromagnetic field is inversely proportional to equilibrium temperature $T'$ as 
   $ \lambda_{max} =
      b/T'$, where $b$ is a constant \cite{planck2013theory}. From eq.(\ref{xi}) the correlation length $\xi_l$ of thermal fluctuations is comparable with wavelength $\lambda_{max}$ of electromagnetic fluctuations and also it can be observed that as $\overline{T} \rightarrow 0$, $\xi_l \rightarrow \infty$. This implies that thermal fluctuations at different points in $d$ dimensional space are independent. Here we assume that statistical description of thermal field having an underlying free field behaviour gives rise to thermal Casimir-like pressure. So taking into account only the free energy term (first term of eq.(22)), we compare
     \begin{equation}
         \left(\frac{E}{V}\right)_C = \frac{ \pi^2}{240 a_\mu^4} \approx \frac{-t_1K^2}{128u^2\xi_l^{4}} = -\frac{t_1}{2}\overline{T}^4 \label{compare} \end{equation}
     where $\left(\frac{E}{V}\right)_C $ is Casimir pressure \cite{Casimir:1948dh} between two perfectly conducting plates and $a_\mu$ is distance between the two plates in natural units which is used henceforth. Assuming $t_1=t'(\overline{T})= \gamma_1$, a constant for $\overline{T}\rightarrow0$ from eq.(\ref{t1}), expression of negative energy density in eq.(\ref{compare}) leads to an attractive force. We identify that force between the plates is pressure due to thermal fluctuations at distances $a_\mu$ comparable with correlation length $\xi_l$. These results are qualitatively in close correspondence with Ref.\cite{Bezerra2011} at low-temperature limit where a scalar field was considered in background of a static universe with topology $S^3 \times R^1$ and $a_\mu$ was related to scale factor in Friedmann universe. Here $\xi_l$ is related to scale factor in cosmological sense. Further we can try to incorporate temperature dependence on this thermal Casimir-like energy using successive corrections from $t$, $u$ and $K$ at higher temperatures. \\ 
     The third term of eq.(\ref{compare}) resembles Stefan-Boltzmann law of radiation \cite{planck2013theory}, which gives radiation pressure as 
     \begin{equation}
          -\frac{\gamma_1}{2}\overline{T}^4. \label{sb} \end{equation}
         Assuming this is true we can also get an expression for thermal Casimir-like force in inertial frames with speed $v_c$ (fraction of light speed) for dynamical system. The expression for inertial relativistic Stefan-Boltzman Law is \cite{2015arXiv150706663L}
         \begin{equation}
             \left[\frac{E}{V}\right]_{v_c}= \frac{1}{15} \left[\frac{E}{V}\right]_0 \left[\frac{(1+v_c)^2 (1-v_c^2)^{3/2} (v_c^4-6 v_c^3+ 15 v_c^2 -20 v_c +15)}{(1-v_c^4)}\right],
         \end{equation}
         where $\left[\frac{E}{V}\right]_{v_c}$ is radiation pressure in inertial frame at speed $v_c$ and $\left[\frac{E}{V}\right]_{0}$ is radiation pressure when $v_c=0$. Similarly at non-relativistic limit ($v_c \ll 1$) thermal Casimir-like force for dynamical field transforms as 
        \begin{equation}
            \left[\left(\frac{E}{V}\right)_C\right]_{v_c}=\frac{-\gamma_1K^2}{128u^2\xi_l^{4}} \left[1+\frac{14}{3}v_c+O(v_c^2)\right]. \label{relcasimir}
        \end{equation}
 \section{Perturbative Wilson's Renormalization Group}
  The prescription being used from beginning is for thermal fields in zero-temperature limit ($\overline{T} \rightarrow 0$, $\xi_l \rightarrow \infty$ from eq. (21)). Casimir energy was derived from free energy between two thick plates by taking limiting case of $a_\mu \rightarrow \infty$ in Lifshitz theory \cite{Lifshitz:1956zz}. In same philosophy we consider domain over which we can use self-similarity or dilation symmetry to perturbatively renormalize thermal fluctuations, where correlations are completely unrelated. Previously in Ref.\cite{Bezerra2011} renormalized Casimir energy density was derived at limit, scale factor $\rightarrow \infty$. \\ We take the entire partition function for most general form of Landau-Ginzburg Hamiltonian \cite{kardar2007statistical} \begin{equation}
    \Phi = \int d^d \mathbf{x} \left[ \frac{t}{2} T^2 (\mathbf{x}) + u T^4 (\mathbf{x}) + v T^6 (\mathbf{x}) + \cdots+ \frac{K}{2}(\nabla T (\mathbf{x}))^2 + \frac{L}{2}(\nabla^2 T (\mathbf{x}))^2 + \cdots\right]
    \end{equation} to be 
  \begin{equation}
  Z= \int DT(\mathbf{x}) \exp[- \Phi_0 -\textit{U} ]  
  \end{equation} such that \begin{multline}
\Phi = \Phi_0 +\textit{U} \equiv \int d^d \mathbf{x} \left[ \frac{t}{2} T^2 (\mathbf{x})+ \frac{K}{2}(\nabla T (\mathbf{x}))^2 + \frac{L}{2}(\nabla^2 T (\mathbf{x}))^2 + \cdots\right]+  \\ u\int d^d \mathbf{x} T^4 (\mathbf{x}) + v\int d^d \mathbf{x} T^6 (\mathbf{x}) +\cdots \ . 
\end{multline} 
Here $\Phi_0$ is exactly solvable considering Gaussian solution and $U$ is the perturbative interaction. Introducing Fourier modes $T(\mathbf{x})=\sum_{\mathbf{q}} \frac{1}{V}T_\mathbf{q} e^{-i\mathbf{q}.\mathbf{x}}$ we perform coarse-graining of renormalization procedure by subdividing fluctuations into two components,
     \begin{equation}
T_\mathbf{q} = \Bigg\{\begin{array}{l}
   \ \widetilde{T_\mathbf{q}} \ \ \hbox{for}\ 0<\mathbf{q}< \Lambda/b \\ 
  \ \sigma_\mathbf{q} \ \  \hbox{for}\ \Lambda/b<\mathbf{q}< \Lambda.
  \end{array}
  \end{equation}
  $\Lambda$ corresponds to shortest possible wavelength of these fluctuations. And $\Lambda/b$ ($b>1$) is mode of the cutoff length at which thermal fields are coarse-grained.  From results in previous section we assume that suppose we take $\Lambda$ to be the mode of smallest length of fluctuations, which influences the plates under thermal Casimir effect and $\Lambda/b$ corresponds to longest fluctuations influencing the plates, we can account for all the factors responsible in thermal Casimir-like effect by studying the probabilistic weight obtained only from $\sigma_\mathbf{q} $ (ignoring $\widetilde{T_\mathbf{q}}$) terms. We obtain the partition function as
    \begin{multline}
  Z= \int d\widetilde{T_\mathbf{q}} d\sigma_\mathbf{q} \exp\left[-\int_{0}^{\Lambda/b}\frac{d^d\mathbf{q}}{(2\pi)^d} |\widetilde{T_\mathbf{q}}|^2\left(\frac{t+K q^2+L q^4 + \cdots}{2}\right) \right] \\ \exp\left[-\int_{\Lambda/b}^{\Lambda}\frac{d^d\mathbf{q}}{(2\pi)^d} |\sigma_\mathbf{q}|^2\left(\frac{t+K q^2+L q^4 + \cdots}{2}\right) \right] \\ \exp\left[-U[\widetilde{T_\mathbf{q}},\sigma_\mathbf{q}]\right].
  \end{multline}
  $\sigma_\mathbf{q}$ modes can be integrated out of the above integral as a Gaussian solution by taking $\left\langle e^{(-U[\widetilde{T_\mathbf{q}},\sigma_\mathbf{q}])} \right\rangle_{\sigma}$ inside the integral, which is a Gaussian of $\sigma$ modes but not of $\widetilde{T}$ modes. 
  \begin{multline}
  \ln Z= \ln\left(\exp\left[-\frac{V}{2}\int_{\Lambda/b}^{\Lambda} \frac{d^d\mathbf{q}}{(2\pi)^d} \ln(t+K q^2+L q^4 + ...)\right]\right) \\ + \ln \left( \int d\widetilde{T_\mathbf{q}} \exp\left[-\int_{0}^{\Lambda/b}\frac{d^d\mathbf{q}}{(2\pi)^d} |\widetilde{T_\mathbf{q}}|^2\left(\frac{t+K q^2+L q^4 + ...}{2}\right) \right] \left\langle e^{(-U[\widetilde{T_\mathbf{q}},\sigma_\mathbf{q}])} \right\rangle_{\sigma}\right).
  \end{multline}
  Expanding the term $\left\langle e^{(-U[\widetilde{T_\mathbf{q}},\sigma_\mathbf{q}])} \right\rangle_{\sigma}$ inside the integral of second term and evaluating the terms contributing in region of $\sigma$ modes, we get corrections to the first significant $\sigma$ modes term in eq. (32) as,
  \begin{equation}
  \ln\left\langle e^{(-U[\widetilde{T_\mathbf{q}},\sigma_\mathbf{q}])} \right\rangle_{\sigma}=-\left\langle U \right\rangle_{\sigma} + ... \ .
  \end{equation}
  Considering only the first order correction term,
  \begin{multline}
  \left\langle U \right\rangle_{\sigma} = u \int \frac{d^d\mathbf{q}_1d^d\mathbf{q}_2d^d\mathbf{q}_3d^d\mathbf{q}_4}{(2\pi)^{4d}}  \delta^d(\mathbf{q}_1+\mathbf{q}_2+\mathbf{q}_3+\mathbf{q}_4) (2\pi)^d \\ \left\langle (\widetilde{T_{\mathbf{q}_1}}+\sigma_{\mathbf{q}_1})(\widetilde{T_{\mathbf{q}_2}}+\sigma_{\mathbf{q}_2})(\widetilde{T_{\mathbf{q}_3}}+\sigma_{\mathbf{q}_3})(\widetilde{T_{\mathbf{q}_4}}+\sigma_{\mathbf{q}_4}) \right\rangle.
  \end{multline}
  Considering only the term contributing in $\sigma$ modes,
   \begin{equation}
  \left\langle U \right\rangle_{\sigma} = u \int \frac{d^d\mathbf{q}_1d^d\mathbf{q}_2d^d\mathbf{q}_3d^d\mathbf{q}_4}{(2\pi)^{4d}} \delta^d(\mathbf{q}_1+\mathbf{q}_2+\mathbf{q}_3+\mathbf{q}_4) (2\pi)^d \left\langle \sigma_{\mathbf{q}_1}\sigma_{\mathbf{q}_2}\sigma_{\mathbf{q}_3}\sigma_{\mathbf{q}_4} \right\rangle.
    \end{equation}
    It is obvious from evaluation of eq. (32) that the leading term contributing to $\sigma$ modes is first term, and the term from eq. (35) gives the first order correction term. We considered thermal Casimir-like energy can be obtained from momentum-shell of only $\sigma_\mathbf{q} $ fluctuations. Hence considering only the leading term contributing to $\sigma$ modes we get energy of described system as 
    \begin{equation}
    \left(\frac{E}{V}\right)_C \propto \frac{\partial \ln Z}{\partial t}= -\frac{1}{2}\int_{\Lambda/b}^{\Lambda} \frac{d^d\mathbf{q}}{(2\pi)^d} \frac{1}{(t+K q^2+L q^4 + \cdots)}.
    \end{equation}
    Taking Fourier modes to be spherically symmetric, and $S_d$ be the solid angle in $d$ dimensions energy density is given by
    \begin{equation}
   \left(\frac{E}{V}\right)_C \propto -\frac{1}{2}\int_{\Lambda/b}^{\Lambda} \frac{d\mathbf{q}}{(2\pi)^d} \frac{S_d\mathbf{q}^{(d-1)}}{(t+K q^2+L q^4 + \cdots)}.
    \end{equation}
    Taking $\mathbf{q}=\sqrt{t/K}x$ and $k_d=S_d/(2\pi)^d$, energy density accounting for thermal casimir-like effect is
      \begin{equation}
    \left(\frac{E}{V}\right)_C \propto -\frac{k_d}{2}\left(\frac{t}{K}\right)^{d/2}\frac{1}{t}\int_{\frac{\Lambda\sqrt{K}}{b\sqrt{t}}}^{\Lambda\sqrt{K/t}}  \frac{x^{(d-1)}dx}{(1+x^2+Lt x^4/K^2 + \cdots)}.
    \end{equation}
    Ignoring higher order terms of $x$ (scaled modes of $\mathbf{q}$) in the denominator and taking $d=4$, energy density can be computed using standard integral,
   \begin{equation}
   \int \frac{x^3dx}{(1+x^2)}=\frac{x^4}{4}-\frac{x^6}{6}+\frac{x^8}{8}+\cdots \ .
   \end{equation}
    Ignoring higher order terms of $x$ in above integral energy density is computed as
    \begin{equation}
   \left(\frac{E}{V}\right)_C \propto -\frac{k_4}{8t}\frac{\Lambda^4}{b^4}[b^4-1] .
    \end{equation}
    We can deduce from above expression that this energy density is equivalent to thermal casimir pressure in eq.(\ref{compare}), for $d=4$ and $b \gtrapprox 1$.  The condition $b \gtrapprox 1$ implies that, only fluctuations ranging close to thin plates are considered. Considering $4$th dimension to be time, it can be interpreted as thermal field coarse-grained over space-time dimensions is used to derive an expression similar to Casimir energy. The coarse-grained field ($\widetilde{T_\mathbf{q}}$) and field before it ($T_\mathbf{q}$) have similar probabilistic weights, when we consider perfect self-similarity condition being implied. This statistically similar weight for coarse-grained field can be obtained by rescaling and renormalization. This implies that thermal Casimir-like effect can be interpreted as scale invariant behaviour of thermal field in Euclidean space-time at limit of absolute zero temperature. 
     \section{Conclusion}
    Considering thermal field as a statistical field, energy density of the field in limit of absolute zero temperature is compared with that from thermal Casimir effect by relating characteristic length scale, $\xi_l$ with scale factor of closed Friedmann universe in a cosmological sense. We have further argued using the coarse-graining procedure of renormalization that this thermal Casimir-like effect can be attributed to scale invariant behaviour of thermal field in $d=4$, critical dimension. This is self sufficient in a manner that the theory is driven towards free field theory at the criticality by Wilson's renormalization group for $d=4$ \cite{Wilson:1973jj}. This study is confined to real scalar field at one limit, in the vicinity of Gaussian fixed point. \\ Further studies can be tried using Halpern-Huang directions for a quantum scalar field \cite{huang1,huang2,Huang_2012} to understand cosmological Casimir effect, which can also explain the cosmological constant \cite{Weinberg:1988cp}. The Halpern-Huang potential have unique form of Kummer function having two singularities \cite{abr65}, which perhaps can correspond to energy density of Casimir effect at one limit and energy density of the cosmological constant at another asymptotic limit.

\bibliographystyle{ieeetr}
\bibliography{sample.bib}

\end{document}